\documentclass[11pt,twoside]{article}

\usepackage{asp2006}
\usepackage{epsf}
\usepackage{lscape}

\newcommand{\spitzer}{{\it Spitzer}}
\newcommand{\mi}{$\mu$m}
\newcommand{\av}{{A$_V$}}
\newcommand{\hii}{H{\sc ii}}

\newcommand{\mm}{M\,33}
\newcommand{\dusty}{{\it DUSTY}}
\newcommand{\tin}{T$_{\rm in}$}

\begin{document}
\title{\spitzer\ photometry of discrete sources in \mm}   
\author{S. Verley$^{1}$, L. K. Hunt$^{2}$, E. Corbelli$^{1}$, C. Giovanardi$^{1}$}   
\affil{$^{1}$INAF-OAA, $^{2}$INAF-IRA}    

\begin{abstract} 
Combining the relative vicinity of the Local Group spiral galaxy \mm\ with the \spitzer\ images, we investigate the properties of infrared (IR) emission sites and assess the reliability of the IR emission as a star formation tracer. We compared the photometric results for several samples of three known types of discrete sources (\hii\ regions, supernovae remnants and planetary nebulae) with theoretical diagnostic diagrams, and derived the spectral energy distribution (from 3.6 to 24~\mi) of each type of object. Moreover, we generated a catalogue of 24~\mi\ sources and inferred their nature from the observed and theoretical colours of the known type sources. We estimated the star formation rate in \mm\ both globally and locally, from the IR emission and from the H$\alpha$ emission line.
\end{abstract}



\section{The IR emission of nebulae in M\,33}
In the four IRAC and 24~\mi\ MIPS bands, we performed photometry of catalogued discrete sources of various type: \hii\ regions, SNRs and PNe. To better understand the IRAC/MIPS colour diagrams of discrete sources, we assembled a series of theoretical models (blackbodies, SB99, Cloudy, \dusty. A PAH-dominated spectrum such as the diffuse ISM and the small dense Cloudy \hii\ region have virtually identical IRAC colours to some of the \tin\,=300\,K \dusty\ models. The implication is that strong PAH features can be easily confused with a rising continuum in the MIR. The only way to resolve the ambiguity is through detailed spectral information {\it or by using the 24\,$\mu$m filter in conjunction with the IRAC 8\,$\mu$m one}. Indeed, only 2 IRAC filters, together with MIPS-24, are sufficient to separate strong continuum emission in the MIR from strong PAH features.

In the colour diagram of the known type sources, we observe a continuity among the various types: the \hii\ regions appear more dusty than the PNe, which behave more like stellar objects. \hii\ regions are better modelled by pure Cloudy \hii\ regions (including PAHs) and by \dusty\ models with inner-shell temperatures around 600~K and low extinction. Statistically, the \hii\ regions are more concentrated near the regions defined by low values of the two shortest IRAC channels and/or equivalently high values of the two longer IRAC wavelengths. The most intense PAH bands (6.2, 7.7, 8.6~\mi) are located in the 5.8 and 8.0~\mi\ channels and this may explain the colours of \hii\ regions.

\section{The 24~\mi\ emission and star formation}
\spitzer\ images provide a unique opportunity to compile vast catalogues of MIR
discrete sources down to stellar luminosities for galaxies in the Local Group: we extracted 515 sources at 24~\mi\ (24Ss). In the colour diagrams, the region with the highest concentration of 24Ss matches well the one of \hii\ regions, although some of the 24Ss extend into the regions populated by SNRs and PNe. However, neither a pure blackbody, Cloudy, or the two \dusty\ models can alone reproduce the colour distribution of the 24Ss. The cumulative luminosity function of the 24Ss displays a double slope behaviour, markedly steeper at the high luminosity tail. The change of slope, in the simplest scenarios, represents the change of regime between poor and rich clusters. The luminosity distribution appears to precisely match the whole range expected for ionised regions and their complexes: from the faintest B2 stars (0.1~mJy) to the brightest clusters containing more than 1000 ionising stars (1.8~Jy). We compare the star formation rates inferred from the MIR emission with that derived from H$\alpha$ emission (uncorrected from extinction). At large scale, the two methods give a SFR of the inner disk (within 5~kpc) of about 0.2~M$_\odot$~yr$^{-1}$. The relation between 24~\mi\ emission and \hii\ regions is also investigated by searching directly for H$\alpha$ counterparts of the 24Ss: the SFR for the single sources in IR and H$\alpha$ are consistent but the scatter is rather large and not imputable to extinction alone.

\section{Conclusions}
The colours of the typical IR emissions of \hii\ regions, supernovae remnants and planetary nebulae are continuous among the different samples, with overlapping regions in the diagnostic diagrams. The comparison between the model results and the colours of \hii\ regions indicates a dusty envelope at relatively high temperatures $\sim$600~K, and moderate extinction \av $\le$ 10. The 24~\mi\ sources IR colours follow the regions observationally defined by the three classes of known objects but the majority of them represent \hii\ regions. The derived total IR luminosity function is in fact very similar to the \hii\ luminosity function observed in the Milky Way and in other late type spirals. Even though our completeness limit is $5\times10^{37}$~erg~s$^{-1}$, in low density regions we are able to detect sources five times fainter than this, corresponding to the faintest possible \hii\ region. The 8 and 24~\mi\ luminosities within the central 5~kpc of \mm\ are comparable and of order $4\times10^{28}$ erg~s$^{-1}$ Hz$^{-1}$ ($\nu L_\nu(8)\,=\,1.5\times10^{42}$ and $\nu L_\nu(24)\,=\,4.4\times10^{41}$~erg~s$^{-1}$). We estimate the total IR emission in the same region of \mm\ to be 10$^9$~L$_\odot$. The discrete sources account for about one third of the 24~\mi\ emission while the rest is diffuse. From the IR emission, we derive a star formation rate for the inner disk equal to 0.2~M$_\odot$~yr$^{-1}$.

\acknowledgements 
We wish to thank the organizing committees for this interesting conference.


\end{document}